\documentclass[a4paper]{report}
\usepackage[utf8]{inputenc}
\usepackage[T1]{fontenc}
\usepackage{microtype}
\usepackage{setspace}
\usepackage{RJournal}
\usepackage{amsmath,amssymb,array}
\usepackage{booktabs}

\begin{document}

\begin{article}
\newcommand{\indep}{\raisebox{0.05em}{\rotatebox[origin=c]{90}{$ \models $}}}
\newcommand\packageName{\CRANpkg{CausalModels}}

\title{An R package for parametric estimation of causal effects}
\author{by Joshua Wolff Anderson and Cyril Rakovski}

\maketitle

\abstract{
  This article explains the usage of R package \href{https://CRAN.R-project.org/package=CausalModels }{\CRANpkg{CausalModels}}, which is publicly available on the Comprehensive R Archive Network. While packages are available for sufficiently estimating causal effects, there lacks a package that provides a collection of structural models using the conventional statistical approach developed by~\cite{hernan2020causal}. \CRANpkg{CausalModels} addresses this deficiency of software in R concerning causal inference by offering tools for methods that account for biases in observational data without requiring extensive statistical knowledge. These methods should not be ignored and may be more appropriate or efficient in solving particular problems. While implementations of these statistical models are distributed among a number of causal packages, \packageName\ introduces a simple and accessible framework for a consistent modeling pipeline among a variety of statistical methods for estimating causal effects in a single R package. It consists of common methods including standardization, IP weighting, G-estimation, outcome regression, instrumental variables and propensity matching.
}

\section{Introduction}
\subsection{Definition of Causality}
Causality has been defined with the identification of the cause or causes of a phenomenon by establishing covariation of cause and effect, a time-order relationship with the cause preceding the effect, and the elimination of plausible alternative causes; see~\citet{shaughnessy2000research}. To claim a specific causal effect between two variables is quite a strong claim. First, there needs to be well-defined treatment and outcome with an established covariance. Second, the treatment must proceed the observed outcome. Third, there must be no other present confounders, i.e., other "treatments" that could have their own causal effect; see~\citet{judea2010introduction}.
\begin{table}[tp]
\centering
\caption{Counterfactuals example with binary treatment and outcome}
\begin{tabular*}{\textwidth}{c |@{\extracolsep{\fill}} ccccc}
\toprule                 & $Z$    & $Y$       & $Y^{z=1}$     & $Y^{z=0}$ \\
\midrule \\
$Person_1$         & 1      & 1         & 1             & -         \\
$Person_2$         & 0      & 1         & -             & 1         \\
$Person_3$         & 1      & 0         & 0             & -         \\
$Person_4$         & 0      & 0         & -             & 0         \\
$Person_5$         & 0      & 1         & -             & 1         \\
$Person_6$         & 1      & 1         & 1             & -         \\
\centering \vdots      & \vdots & \vdots    & \vdots        & \vdots    \\
$Person_n$         & 0      & 0         & -             & 0         \\       
\bottomrule 
\end{tabular*}
\label{counterfactualTable}
\end{table}

While these conditions are not perfect parameters for inferring a causal relationship between a treatment and outcome, they help researchers remove strong bias from their studies; see~\citet{hammerton2021causal}. A causal effect found in a causal inference study is almost never the true causal effect, rather a less-biased estimate that is significantly closer to the true causal effect of the treatment on the outcome. To calculate a true causal effect would require "counterfactual" outcomes that cannot be measured; see~\citet{judea2010introduction}.

To describe a counterfactual outcome, let us define some treatment $Z$ and an outcome $Y$. Using $z$ to notate a given value of $Z$, we can write a counterfactual outcome as $Y^z$. A counterfactual describes the hypothetical outcome if the population was given a particular treatment. For example, if $z \in [0, 1]$, i.e. $Z$ is a binary treatment, then the hypothetical (counterfactual) outcome would be $Y^{z=1}$ if the entire population were treated. The counterfactual outcome if the population was not treated would be $Y^{z=0}$. Table~\ref{counterfactualTable} has example data showing the relationship between our observed variables and the known counterfactual values. We can use several methods to compare counterfactual outcomes to calculate the causal effect on different scales such as on the additive (\ref{diff}) and the multiplicative (\ref{ratio}) scales; see~\citet{sjolander2021novel} and~\cite{vanderweele2015explanation}.

Unfortunately, we cannot measure a particular individual's outcome given all values of treatment since we cannot independently give multiple treatments to one individual at the same exact point in time. This means we cannot measure all counterfactuals for every individual as we only know the outcome for the treatment they were given. To address this, we make assumptions based on expert knowledge about the causal relationships in the data we can observe. When these assumptions hold, the adjusted associative effect is equal to the causal effect. The general required conditions are called identification assumptions; see~\citet{hernan2020causal} and the causal structure is usually represented graphically with Directed Acyclic Graphs (DAGs); see~\citet{judea2010introduction}.

\subsection{Identification Assumptions}
There are three main assumptions made to infer a causal effect from association: exchangeability, positivity, and consistency. Exchangeability is the assumption that $Y^z \indep Z$ or all counterfactual outcomes are independent of treatment; see~\citet{hernan2006estimating}. This means that with a binary treatment, the individuals who were treated would have the same average outcome as the untreated had they not been treated, and the individuals who were not treated would have had the same average outcome as the treated if they were treated; see~\citet{judea2010introduction}. Usually, this is not the case, but we have an alternative, conditional exchangeability. This is the same assumption but made in subpopulations of our data. If we have some covariates $X$ by which we can stratify the data, it is more common to find $Y^z \indep Z | X$ or all counterfactual outcomes are independent of treatment conditional on values in $X$; see~\citet{hernan2020causal}. In this case, we can find the causal effect within each strata and find the average causal effect for the population by taking the weighted average of the effect in each strata (standardization).

Positivity is the second assumption of causal inference which is relevant under conditional exchangeability. The assumption is that there are individuals both assigned and not assigned treatment within each combination of values of $X$ in the study. This is an overlooked, but essential part of causal inference; see~\citet{westreich2010invited}. Positivity is not always guaranteed especially as $X$ becomes larger.

Consistency is the third assumption which requires well-defined treatments to be present in the data. In other words, consistency may not hold if $z \not \in Z$ for any $Y^z$ or if our study does not acknowledge all values for $Z$. Possible treatments and observed interventions are not always the same; see~\citet{rehkopf2016consistency}. When we claim a causal effect, we explicitly make these three assumptions about the data.

\subsection{Randomized Experiments}

Even With randomized experiments, we still have missing values for all but one of the counterfactuals in any study. Although in the context of a randomized experiment, the missing values of the counterfactuals happen by chance. This is extremely valuable as randomized experiments are expected to produce exchangeability due to data missing completely at random (MCAR). This means that we do not need to be aware of or concerned about bias from outside effects or measure any covariates $X$; see~\citet{hernan2020causal}.

Consider Table~\ref{counterfactualTable} again. If we first examine when $z=0$, then for our assumption to hold, P($Y^{z=0}|Z = 1$) = P($Y^{z=0}|Z = 0$) = P($Y^{z=0}$) must be true. Additionally for $z=1$, P($Y^{z=1}|Z = 1$) = P($Y^{z=1}|Z = 0$) = P($Y^{z=1}$) must be true; see~\citet{judea2010introduction}. Since we do not have all the values for a given $Y^z$, we cannot actually make these calculations, making it difficult to assume exchangeability outside of a randomized experiment.

Sometimes an experiment may not have randomized treatment, but still randomly assigned treatment within specific groups. This is called a conditionally randomized experiment; see~\citet{hernan2020causal}. If $Y^z \indep Z$ does not hold, many times we may find or design an experiment where $Y^z \indep Z | X$ for some covariates $X$. If within each of our levels of $X$ treatment is randomly assigned, we can treat each strata as conditionally exchangeable. Because any given $X$ could have imbalance, there are several methods to adjust accordingly; see~\citet{hennessy2016conditional, hernan2020causal, judea2010introduction}. Methods discussed later consist of a weighted average in calculating the population causal effect from averaging the effect in each strata.

In reality, the vast majority of the available data are observational. Typically, these do not involve random treatment, and therefore, we cannot immediately assume exchangeability, positivity or consistency. In randomized experiments, these hold by design. In observational studies, these have to be assumed and carefully examined; see~\citet{hernan2020causal}.

\subsection{Observational Studies}

Unfortunately, since most data arise from observational studies, methods for causal inference are needed for observational data. This means that there must be adjustment for biases that occur when there is a lack of randomized assignment of treatment. Typically, these are done with matching, stratification, or covariate adjustments; see~\citet{rosenbaum2005observational}.

There are three main types of biases that occur in observational data that are relevant for causal inference: selection bias, measurement bias, and confounding; see~\citet{hernan2020causal}. Selection bias describes an array of biases that occur in observational studies such as inappropriate control selection, differing rates of loss-to-follow up, volunteer bias, etc.; see~\citet{arnold2016brief, hernan2004structural}. Measurement bias adds or removes association between the treatment and outcome depending on how the data was measured, which usually includes self-reporting surveys and error in measurement; see~\citet{arnold2016brief, hernan2020causal}. Confounding is bias that is introduced into the effect of the treatment of interest due to a common causal effect on the outcome; see~\citet{van2010confounding}. 

While these biases can be adjusted for nonparametrically in some simple scenarios, \packageName\ is primarily concerned with the parametric solutions using structural models; see~\citet{judea2010introduction}. When adjustment for bias in observational studies has been made, it can be said that P($Y^{z=1} = 1$) = P($Y = 1 | Z = 1$) and P($Y^{z=0} = 1$) = P($Y = 1 | Z = 0$) in the case of a binary outcome and treatment under the identification assumptions. The same property applies for $Y^{z},\ Y = 0$; see~\citet{hernan2020causal}.

\section{Related Work}

While causal inference has been present for a number of years in fields like economics and medicine, the rise of machine learning and strong predictive models has brought a recent revival in interest of causal inference in computer science; see~\citet{blakely2020reflection}. This has led to a number of open-source causal modeling packages in R.

Several packages in R have already developed methods similar to those contained in \packageName\ separately. \CRANpkg{causalweight} is a package developed for using IP weighting to calculate several causal analyses using binary treatment; see~\citet{bodory2018causalweight}. \CRANpkg{WeightIt} allows for modeling propensity scores in R; see~\citet{greifer2020guide}. \CRANpkg{drtmle} developed at Emory University provides a function to calculate estimates for counterfactual values using doubly-robust estimator; see~\citet{benkeser2017drtmle}. \CRANpkg{DTRreg} provides methods for G-estimation for causal effects; see~\citet{wallace2017r}. There are also a number of packages in R that provide similar individual methods. 

It should be noted there is a package \CRANpkg{causaleffect} that provides methods for solving similar problems described in this paper; see~\citet{tikka2018identifying}. While this well-known package is sufficient for causal inferences, it is not exhaustive as it focuses on methodologies of do-calculus by Pearl. \CRANpkg{CausalModels} uses a more traditional statistical approach developed by Robins which provides an alternative philosophy in the structure of causal modeling; see~\citet{judea2010introduction} and~\cite{hernan2006estimating}.

\section{Nonparametric methods}

There are fundamental, nonparametric methods that allow for calculation of causal effects under the assumption of a randomized experiment. This means that treatment is randomly assigned and is independent of the outcome with no biases or confounding. The methods include the difference, ratio, and odds ratio of average treatment effects; see~\citet{hernan2020causal} shown in equations \ref{diff}-\ref{or}. For continuous values, the terms consist of expected values rather than probabilities. 

\begin{equation}\label{diff}
    P(Y^{z=1}=1) - P(Y^{z=0}=1)
\end{equation}

\begin{equation}\label{ratio}
    \frac{P(Y^{z=1}=1)}{P(Y^{z=0}=1)}
\end{equation}

\begin{equation}\label{or}
    \frac{P(Y^{z=1}=1) / P(Y^{z=1}=0)}
         {P(Y^{z=0}=1) / P(Y^{z=0}=0)}
\end{equation}

In the case of a conditionally randomized experiment, there are two common nonparametric methods, standardization and IP weighting, which can adjust the treatment effect within each strata for an estimate of the population average causal effect. Table~\ref{std_ip} shows the calculations for estimating counterfactuals of a binary treatment and outcome with some matrix of confounders using these two methods. Additionally for IP weighting, it is common to replace the inverse probability, $1/PMF(Z|X)$, with standardized weights using $PMF(Z) / PMF(Z|X)$ instead; see~\citet{hernan2020causal}. When $Z$ is continuous, we would use the PDF instead of PMF. It has been proven that standardization and IP weighting are mathematically equivalent estimators; see~\citet{hernan2006estimating}. Since most observational data is too large to contain samples within all possible values of $X$, \packageName\ provides the parametric versions of standardization and IP weighting.

\begin{table}[tp]
    \centering
    \caption{Estimation of $Y^{z=1}$ and $Y^{z=0}$ using standardization and IP weighting}
    \small
    \begin{tabular}{p{2cm}|p{5.25cm} p{5.25cm}}
    \toprule
    Estimand & Standardization & IP Weighting \\
    \midrule \\\
    \raggedright 
    $P(Y^{z=1}=1)$ & $\underset{x\in X}{\sum} P(Y = 1 \text{\textbar} Z = 1, X = x)P(X=x)$ & $\underset{x\in X}{\sum} P(Y = 1 \text{\textbar} Z = 1, X = x) \frac{1}{PMF(Z\text{\textbar} X)}$\\ 
    \\ \midrule \\
    \raggedright 
    $P(Y^{z=0}=1)$ & $\underset{x\in X}{\sum} P(Y = 1 \text{\textbar} Z = 0, X = x)P(X=x)$ & $\underset{x\in X}{\sum} P(Y = 1 \text{\textbar} Z = 0, X = x) \frac{1}{PMF(Z\text{\textbar} X)}$\\ \\
    \bottomrule
    \end{tabular}
    \label{std_ip}
\end{table}

\section{Structural Models}

For all parametric methods in \packageName, a binary treatment and continuous outcome are assumed. Covariates can be of any type valid for training a statistical model. Support for less restricted variables may be added in the future.

A number of the methods in \packageName\ use structural models. Structural models are used in causal inference as a parametric method of adjusting for biases in observational studies; see~\citet{robins2000marginal}. Since in many cases we cannot measure a conditional mean of the outcome in the treated and the untreated due to high dimensional data, which does not contain all possible samples, we use modeling to make parametric estimates for inference. Consider the following linear model~(\ref{linear}):

\begin{equation}\label{linear}
    E(Y | Z, X) = \beta_0 + \beta_1 Z + \beta_2 X
\end{equation}

Using structural models, we can obtain the conditional mean with $\widehat{E}$($Y | Z, X$) through finding $\hat{\beta_0},\ \hat\beta_1$, and $\hat\beta_2$ ; see~\citet{hernan2020causal}.

\section{The underlying algorithms}

\subsection{Standardization}

The first family of models that are structural models in the package are parametric variations of standardization and IP weighting. To restate, a large matrix of confounders usually means that it is impossible to calculate unbiased treatment effects nonparametrically as many strata are not represented in real world data. 

Consider again the formula for standardization in Table \ref{std_ip} in the case of a continuous outcome, $\underset{x\in X}{\sum} E(Y \text{\textbar} Z = z, X = x)P(X=x)$. Rather than estimating $P(X=x)$ we can use the definition of expectation to change the sum to a nested expectation, $E(E(Y|Z=z, X=x))$ since this represents the weighted mean. 

\begin{align}
    \begin{split}\label{std_model}
        \widehat{E}(Y^{z}) &= \underset{x\in X}{\sum} E(Y \text{\textbar} Z = z, X = x)P(X=x) \\
        &= E(E(Y|Z=z, X=x)) \\
        &= \frac{1}{n}\sum_{i=0}^{n} \widehat{E}(Y \text{\textbar} Z = z, X_i) \\
        &= \beta_{0} + \beta _{1} Z + \beta_{2} Z X + \beta_{3} X \\
    \end{split}
\end{align}

To estimate this value, we can use $\frac{1}{n}\sum\limits_{i=0}^n \widehat{E}(Y \text{\textbar} Z = z, X_i)$ since this double expectation is simply a mean of means; see~\citet{hernan2020causal}. An OLS model can properly estimate this mean of means using the following structural model with continuous outcome and binary treatment (\ref{std_model}).

This form of standardization estimates the population average causal effect on the additive scale; see~\citet{hernan2006estimating}. For this model to provide an unbiased estimator of the population average causal effect, the identification assumptions must hold, the dataset must include all known confounders in $X$ and the model must be correctly specified.

\subsection{IP Weighting}

Similar to standardization, IP weighting can be done parametrically. Rather than using double expectation, we can estimate both terms. For a binary treatment, a logistic regression model (\ref{logit_mod}) on the treatment can used $PMF(Z|X)$ to calculate the inverse probability:

\begin{equation}\label{logit_mod}
    PMF(Z|X) = \frac{1}{1 + e^{-(\beta_0 + \beta_1 X)}}
\end{equation}

This estimate of $PMF(Z|X)$ is referred to as a propensity score; see~\citet{rubin1996matching}. It can be used to calculate the weights by either using $1/PMF(Z|X)$ or creating another logistic model of $Z$ with only an intercept, $\beta_0$, to give $PMF(Z)$ for calculating the standardized weight, $PMF(Z) / PMF(Z|X)$; see~\citet{hernan2020causal}. 

Next, the weights are used in a weighted linear regression of just the treatment and outcome (\ref{wlr_mod}). These weights correctly adjust the regression line for any bias represented in $X$ without including them in the linear model; see~\citet{robins2000marginal}.

\begin{equation}\label{wlr_mod}
    E(Y^z) = \beta_0 + \beta_1 Z
\end{equation}

The output of this model is not the population average causal effect, but rather an estimate for the counterfactual. These types of models are called marginal structural models; see~\citet{hernan2006estimating}. The population average causal effect is actually seen in $\hat \beta_1$, which from the weighted linear regression can be directly interpreted as an estimate for $E(Y^{z=1}) - E(Y^{z=0})$; see~\citet{robins2000marginal}.

\subsection{Outcome Regression}

Outcome regression is the simplest model in the package. It looks very similar to standardization, but rather than having a marginal structural model estimating $\widehat{E}(Y^z)$, it estimates causal effect conditional on $X$; see~\citet{robins2000marginal}:

\begin{equation}\label{outcome_reg}
    \widehat{E}(Y^z | X) = \beta_0 + \beta_1 Z + \beta_2 Z X + \beta_3 X
\end{equation}

For simplicity, the package will initially support linear regression for continuous outcomes. In the future, it may also support more advanced regression methods for different types of outcomes. The model outputs change in average causal effect within each strata for both continuous and binary covariates. For continuous covariates, specific values of interest need to be specified to be displayed in the contrast matrix.

\subsection{Propensity Matching}

Propensity matching is another modeling technique offered in \packageName\ that provides an alternative approach to structural modeling for conditional average causal effects. Consider again the model for the propensity scores, $S(X) = \widehat{E}(Z=1 | X)$. We can adjust for biases in the data without conditioning on $X$ by conditioning on $S(X)$; see~\citet{caliendo2008some}. To account for $S(X)$ representing the probability of treatment and therefore is a continuous value between 1 and 0, a range of propensity scores is required to represent each strata since it is highly unlikely for two individuals to have the same score; see~\citet{hernan2020causal}. In other words, propensity matching can calculate the conditional average causal effects on the additive scale with the following:

\begin{equation}
    E(Y | Z = 1, S(X) = s) - E(Y | Z = 0, S(X) = s)
\end{equation}

\subsection{G-Estimation}

G-estimation is a much more computationally heavy, but the most useful approach to estimating causal effects. If we consider the mean of differences as a structural model, $E(Y^z - Y^{z=0}) = \beta_0 Z$, then we can rearrange the two unknowns, and through consistency, see $Y^{z=0}$ as one unknown (\ref{untreated_gest}); see~\citet{hernan2020causal}. Using the knowledge that $Y^{z=0} \indep Z$, the propensity model dependent on $Y^{z=0}$ (\ref{logit_gest}) should have a coefficient of 0 for $\alpha_1$. After substituting equation \ref{untreated_gest} in equation \ref{logit_gest}, a grid search on $\beta$ will approximate when $\alpha_1$ is zero; see~\citet{robins2000marginal}.

\begin{equation}\label{untreated_gest}
    Y^{z=0} = Y - \beta Z
\end{equation}

\begin{equation}\label{logit_gest}
    P(Z = 1 | Y^{z=0}) = \frac{1}{1 + e^{-(\alpha_0 + \alpha_1 Y^{z=0})}}
    = \frac{1}{1 + e^{-(\alpha_0 + \alpha_1 (Y - \beta Z))}}
\end{equation}

Having an estimate from the logistic model for $Y^{z=0}$ then allows for calculation of the causal effect using estimates for both $Y^z$ and $Y^{z=0}$. In the case where the user may want to estimate $E(Y^{z} - Y^{z=0} | X)$, the problem becomes much more complex as the grid search is high dimensional. This can be seen by the new observed equation for $Y^{z=0}$ (\ref{gest_2d}) where the grid search on the logistic model needs to find values for $\beta_0$ and $\beta_1$ that result in $\alpha_1$ being zero; see~\citet{hernan2020causal}. even though there are close form solutions for the estimation of causal effects via g-estimation is simple settings, in general, the approach is very computational heavy with complexity that grows quickly as $X$ becomes large because $\beta_1$ becomes large as well. Since this method is the most difficult mentioned, only a single parameter estimation of $\beta$ is implemented in both closed and grid search form. Multi-parameter estimation for many betas will be available in future versions.

\begin{equation}\label{gest_2d}
    Y^{z=0} = Y - (\beta_0 Z + \beta_1 X)
\end{equation}

\subsection{Doubly-Robust Estimator}

While the estimates for standardization and IP weighting are equivalent, they are usually not exactly equal. Since they used different methods for calculating weights for adjustments of treatment effects within each strata, the reduction of bias may differ in the two models on the same data, although it is usually unknown which model performs better; see~\citet{hernan2020causal}. For this reason, doubly-robust estimator are used to find a compromise between standardization and IP weighting. Usually, at least one of the two are strong estimators of the true average causal effect. By using both models (\ref{dr_model}), a better, less biased estimate can be found without the knowledge of which model is correctly specified for the problem; see~\citet{funk2011doubly}. 

For $E(Y^{z=1})$, let $s(X) = E(Y \text{\textbar} Z = 1, X)$ and $w(X) = P(Z=1\text{\textbar} X)$ where $E(s(X))$ denotes standardization and $E(\frac{Z Y}{w(X)})$ denotes IP weighting; see~\citet{hernan2020causal}. We can represent the doubly-robust estimator as:

\begin{equation}\label{dr_model}
    \widehat{E}(Y^{z=1}) = \frac{1}{n} \sum_{i=1}^n\left(\hat{s}(X_i) -  
    \frac{Z_i(Y_i - \hat{s}(X))}{\hat{w}(X)} \right)
\end{equation}

There are many other variations of doubly-robust estimators that are valid estimators where $s(X)$ represents an estimate from some outcome model, and $w(x)$ represents some estimate from a treatment model; see~\citet{funk2011doubly}. The package will support any outcome and treatment model that use the R generic predict function. The predictions of the outcome model should equal $\widehat{E}(Y)$, and the predictions from the treatment model should equal $\widehat{P}(Z=1)$.

\subsection{Instrumental Variables}

Instrumental variables is a very different and popular approach for estimating average causal effects. The method is much simpler in that it usually uses a single instrumental variable, $V$, that under four identifying conditions can adjust for bias without controlling for the effect of confounders, $X$. This is a very popular method in the social sciences like economics because of this; see~\citet{angrist1996identification}. In order to adjust for the bias using $V$ instead of $X$, there are an additional set of assumptions that need to be made about the data. There are three main assumptions: $V$ is correlated with $Z$, $V$ and $Y$ do not have a common cause, and $V$ does not affect $Y$ except for the fact that it affects $Z$. These three conditions only allow an interval estimation of the causal effect. These intervals are usually wide and include the null. Thus, we need an additional fourth assumption. There are two variations of the fourth condition: homogeneity and monotonicity. The allow a precise estimation of the causal effect of treatment within specific stratum of the population (treated, compliers). Therefore, we need a fifth condition, no effect modification by the variables mentioned above to extend the stratum-specific causal effect to the entire population. Even with the most lenient interpretation of homogeneity, it is not very likely to hold; see~\citet{fang2012apples}. A more likely fourth assumption used is monotonicity. In short, this assumes there are no individuals in the population who defied their assignment of treatment. In other words, this would include if the individual was assigned to be treated, they did not take the treatment, and if they were not assigned treatment, they took the treatment; see~\citet{hernan2020causal}.

If these assumptions hold, then the estimand in the package for instrumental variables method estimates the average causal effect on the additive scale; see~\citet{angrist1996identification}:

\begin{equation}
    \widehat{E}(Y^{z=1} - Y^{z=0}) = \frac{E(Y | V=1) - E(Y | V=0)}{E(Z|V=1) - E(Z|V=0)}
\end{equation}

\section{Standard Usage of CausalModels}

\subsection{Application in smoking habits}

Model specification is crucial in identifying unbiased effects. We will compare all the models available in the package on National Health and Nutrition Examination Survey (NHEFS) data used in~\citet{hernan2020causal} and was provided by the \CRANpkg{causaldata} package; see \citet{causaldata}. The NHEFS data came from a follow-up study tracking various dimensions of an individual's personal health. 1566 cigarette smokers between 25 and 74 years of age were measured in an initial measurement and follow-up about 10 years later. In this example, we want to estimate the causal effect of quitting smoking on weight. The difference of weight of the follow-up to the baseline is considered the outcome and the act of quitting smoking during the study is considered the treatment. Known confounders are sex, race, age, education level, intensity of smoking, years of smoking, exercise, physical activity in daily life, and initial weight.

\subsection{Initializing package parameters}

A unique framework in the package is introduced through the \verb|init_params()| function. Before the models in the package can be used, the user must specify a continuous outcome, dichotomous treatment, covariates for adjustment, and the dataset. This will globally set these variables for each model by autogenerating a default formula for the user. While these defaults are required to be set, they may be overwritten when fitting a model. This is intended to simplify the modeling process while encouraging consistency of model specification across multiple models. Correct model specification is paramount when estimating causal effects. The default formula can include or exclude interactions and squared terms by using the option \verb|simple|. This function has the following structure:

\begin{example}
init_params(outcome, treatment, covariates, data, simple = FALSE)
\end{example}

Once the function has been run, it will output each of the variables assigned as well as the default formulas for outcome and propensity models respectively.

\begin{example}
R> library(CausalModels)
R> library(causaldata)
R> data(nhefs)
R> nhefs$qsmk <- as.factor(nhefs$qsmk)
R> confounders <- c("sex", "race", "age", "education", "smokeintensity",
  +                 "smokeyrs", "exercise", "active", "wt71")
R> init_params(outcome = wt82_71, 
  +            treatment = qsmk, 
  +            covariates = confounders, 
  +            data = nhefs)
\end{example}
\begin{example}
Successfully initialized!

Summary:

Outcome - wt82_71 
Treatment - qsmk 
Covariates - [ sex, race, age, education, smokeintensity,
               smokeyrs, exercise, active, wt71 ] 

Size - 1566 x 67 

Default formula for outcome models: 
wt82_71 ~ qsmk + sex + race + education + exercise + active + age +
(qsmk * age) + I(age * age) + smokeintensity + (qsmk * smokeintensity) +
I(smokeintensity * smokeintensity) + smokeyrs + (qsmk * smokeyrs) + 
I(smokeyrs * smokeyrs) + wt71 + (qsmk * wt71) + I(wt71 * wt71) 

Default formula for propensity models: 
qsmk ~ sex + race + education + exercise + active + age + I(age * age) +
smokeintensity + I(smokeintensity * smokeintensity) + smokeyrs +
I(smokeyrs * smokeyrs) + wt71 + I(wt71 * wt71)
\end{example}

\subsection{Standardization}

Like most of the methods in the package, the standardization function uses generalized linear models; see~\citet{R_cite}. The function has the following structure:

\begin{example}
standardization(data, f = NA, family = gaussian(), simple = pkg.env$simple, 
    n.boot = 50, ...)
\end{example}

By default, the \verb|standardization()| function will use the formula from the initialize function and use the \verb|glm()| function with where the family is \verb|gaussian()|. The function also uses bootstrapping to generate a confidence interval around the average treatment effect statistic. The \verb|n.boot| parameter determines how many iterations of bootstrapping will run and is set to fifty by default. Printing the model will print the underlying OLS model as well as the estimated average treatment effect.

\begin{example}
R> std_model <- standardization(nhefs, n.boot = 100, simple = T)
R> print(std_model)
\end{example}
\begin{example}
Call:  glm(formula = wt82_71 ~ qsmk + sex + race + education + exercise + 
    active + age + smokeintensity + smokeyrs + wt71, family = family, 
    data = data)

Coefficients:
   (Intercept)           qsmk1            sex1           race1      education2  
      16.09035         3.38117        -1.42930         0.62743         1.02924  
    education3      education4      education5       exercise1       exercise2  
       0.82417         1.45632        -0.04048         0.38389         0.47487  
       active1         active2             age  smokeintensity        smokeyrs  
      -1.11474        -0.43025        -0.20060         0.02073         0.05159  
          wt71  
      -0.09980  

Degrees of Freedom: 1565 Total (i.e. Null);  1550 Residual
  (3132 observations deleted due to missingness)
Null Deviance:	    97180 
Residual Deviance: 85140 	AIC: 10740

Average treatment effect of qsmk:
Estimate -  3.381171 
SE       -  0.4544547 
95
\end{example}

Other generic functions in~R, \verb|summary()| and \verb|predict()| will work and interact in a similar fashion with the underlying OLS model. This function calculates an estimate of the average treatment effect by making copies of the dataset with a reassigned treatment 0 and 1 to make estimates for each counterfactual value. For simplicity, the output of these copies will be called $D_0$ and $D_1$ respectively. This is done by predicting the output for each of the copies and taking the average. The estimate in the output is calculated by taking the difference of the average predicted outcome in each, $\hat{E}(D_1) - \hat{E}(D_0)$. The average value of the predictions, $\hat{E}(D_0)$ and $\hat{E}(D_1)$, are also included in the model object and can be used to manually calculate a statistic alternative to risk difference. 

\begin{figure}[ht]
\centering
\includegraphics[width=300pt]{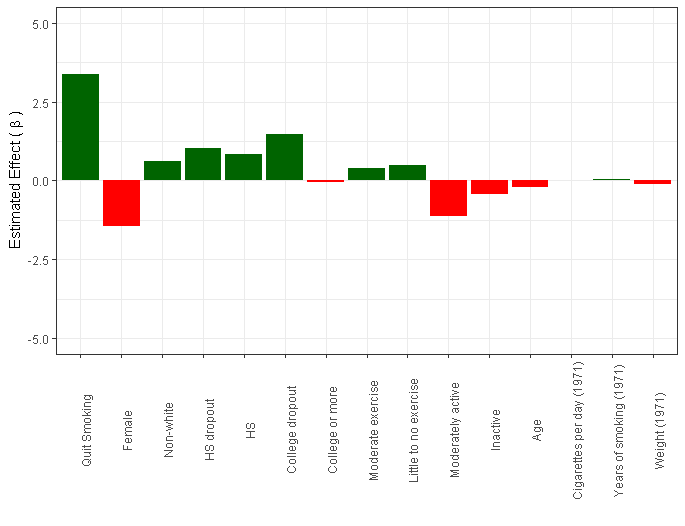}
\caption{Standardization model coefficients representing the estimated effect. The effect for "Quit Smoking" is the estimate for the causal effect of treatment. All other values represent effect modification and adjustment for confounding bias.}
\label{fig:1}
\end{figure}

We can visualizes the effect sizes of the treatment and confounding variables; see Figure~\ref{fig:1}. The model estimates quitting smoking will cause weight loss of approximately 3.38 kilograms. While this model theoretically produces an unbiased estimate of the effect of quitting smoking on weight, there is uncertainty if the model is correctly specified for the problem. For this reason, we will see conflicting estimates among different models. For this reason, we generate an estimate for each model in the package for comparison.

\subsection{IP Weighting}

The \verb|ipweighting()| function produces a similar output but using the inverse probability methodology instead. It has the following structure:

\begin{example}
ipweighting(data, f = NA, family = gaussian(), p.f = NA, 
  p.simple = pkg.env$simple, p.family = binomial(), p.scores = NA, 
  SW = TRUE, n.boot = 0, ...)
\end{example}

The parameters are very similar to~\verb|standardization()| except there is no option to override \verb|simple| since the final model is weighted least squares (WLS) of simply treatment and outcome. The same parameters are available for the underlying propensity model and are given a prefix. For example, to pass a specific formula to the propensity model, the \verb|p.f| parameter should be used and to change the family of the propensity model, the \verb|p.family| parameter should be used. This model can also be called separately and has the following structure:

\begin{example}
propensity_scores(data, f = NA, simple = pkg.env$simple, 
  family = binomial(), ...)
\end{example}

The propensity model will fit the data to predict the probability of treatment using the covariates. The IP weighting model will fit a WLS model with the inverse of the propensity scores. By default, standardized weights are used but this can be changed by setting~\verb|SW| to~\verb|FALSE| If the user wants another propensity model not provided in the package, the propensity scores can be manually input into the function via the~\verb|p.scores| parameter. This parameter will override the modeling of the propensity scores and use the ones given by the user.

\begin{example}
R> ip.model <- ipweighting(nhefs, p.simple = TRUE)
R> print(ip.model)
\end{example}
\begin{example}
Call:  glm(formula = wt82_71 ~ qsmk, family = family, data = data, 
            weights = weights)

Coefficients:
(Intercept)        qsmk1  
      2.553        2.569  

Degrees of Freedom: 1565 Total (i.e. Null);  1564 Residual
Null Deviance:	    115000 
Residual Deviance: 112900 	AIC: 11010

Average treatment effect of qsmk:
Estimate -  2.56858 
SE       -  0.4779627 
95
\end{example}

The coefficient of the WLS model is used as the estimate of the average treatment effect. For this reason, \verb|n.boot| is set to 0, and the standard error calculated by \verb|glm()| is used instead. If \verb|n.boot| is set to a positive value, the standard error calculated from the bootstrapping will be used instead. In the example above, \verb|p.simple| is set to \verb|TRUE| which means the underlying propensity model will use the autogenerated formula without interactions.

\subsection{Outcome Regression}

The function \verb|outcome_regression()| builds a linear model using all covariates. The treatment effects are stratified within the levels of the covariates. The model will automatically provide all discrete covariates in a contrast matrix. This is done using the \verb|glht()| function from the \CRANpkg{multcomp} package; see~\cite{Hothorn2008-zk}. To view estimated change in treatment effect from continuous variables, a list called contrasts, needs to be given with specific values to estimate. A vector of values can be given for any particular continuous variable. The function has the structure:

\begin{example}
outcome_regression(data, f = NA, simple = pkg.env$simple, family = gaussian(), 
  contrasts = list(), ...)
\end{example}

The parameters are similar to \verb|standardization()|. Additionally, there is another parameter \verb|contrasts| which specifies the values of any continuous variables to include in the analysis. This should be a named list corresponding to the continuous variables in the dataset.

\begin{example}
R> outreg.mod <- outcome_regression(data = nhefs,
+                                   contrasts = list(age = c(21, 30, 40),
+                                                smokeintensity = c(5, 20)))
R> print(outreg.mod)
\end{example}
\begin{example}
Average treatment effect of qsmk:

                                         Estimate Std. Error     2.5 
Effect of qsmk at qsmk1                 0.5509460   2.822912 -4.981860 6.083752
Effect of qsmk at sex1                 -0.8862384   2.868126 -6.507662 4.735185
Effect of qsmk at race1                 1.1377836   2.872922 -4.493039 6.768607
Effect of qsmk at education2            1.3684229   2.862462 -4.241899 6.978745
Effect of qsmk at education3            1.1333579   2.846699 -4.446070 6.712786
Effect of qsmk at education4            2.0750350   2.890966 -3.591154 7.741224
Effect of qsmk at education5            0.3717038   2.855788 -5.225537 5.968945
Effect of qsmk at exercise1             0.8573187   2.854214 -4.736839 6.451476
Effect of qsmk at exercise2             0.9060249   2.869506 -4.718103 6.530153
Effect of qsmk at active1              -0.3951223   2.844877 -5.970979 5.180734
Effect of qsmk at age of 21             0.8095362   2.240964 -3.582673 5.201746
Effect of qsmk at age of 30             0.9203606   2.229701 -3.449772 5.290494
Effect of qsmk at age of 40             1.0434988   2.401860 -3.664059 5.751057
Effect of qsmk at smokeintensity of 5   0.7749600   2.786389 -4.686262 6.236182
Effect of qsmk at smokeintensity of 20  1.4470019   2.745636 -3.934346 6.828349
\end{example}

The above example stratifies by all discrete variables and five specific values of continuous variables. The standard error and confidence interval for each strata are included. A p-value for each strata can also be seen by using \verb|summary()| on the model.

\subsection{Propensity Matching}

The function \verb|propensity_matching()| uses either stratification or standardization to model an outcome conditional on the propensity scores. In stratification, the model will break the propensity scores into groups and output a \verb|glht()| object based off a contrast matrix which estimates the change in average causal effect within groups of propensity scores. This is a form of outcome regression. In standardization, the model will output a \verb|standardization()| object that conditions on the propensity strata rather than the covariates. The model can also predict the expected outcome. It has the following structure:

\begin{example}
propensity_matching(data, f = NA, simple = pkg.env$simple, p.scores = NA, 
  p.simple = pkg.env$simple, type = "strata", grp.width = 0.1,
  quant = TRUE, ...)
\end{example}

The parameters are similar to \verb|ipweighting()| for both the outcome and propensity elements of the model. Additionally, there is a parameter \verb|strata| which determines whether the model uses standardization or stratification via outcome regression. To determine how the strata of propensity scores are made, the user can specify to use quantiles with the \verb|quant| parameter. If it is set to \verb|FALSE| then \verb|grp.width| will be used instead to make groups with some width between 0 and 1. It should be noted there may be issues with positivity in the groups with smaller widths. All strata of propensity scores must have a non-zero number of samples in order for the model to work properly. If standardization is used, then the propensity scores are modeled as a continuous value, and unless a specific formula is given, the squared term along with the interaction between propensity scores and treatment are included.

\begin{example}
R> pm.model <- propensity_matching(nhefs.nmv, type = "strata")
R> print(pm.model)
\end{example}
\begin{example}
Average treatment effect of qsmk:

                                             Estimate Std. Error
Effect of qsmk for p.score in [0.051,0.124] 0.2073431   2.394879
Effect of qsmk for p.score in (0.124,0.16]  4.9139249   1.728781
Effect of qsmk for p.score in (0.16,0.189]  4.6981335   1.598030
Effect of qsmk for p.score in (0.189,0.212] 2.2821425   1.454161
Effect of qsmk for p.score in (0.212,0.237] 3.8269831   1.441765
Effect of qsmk for p.score in (0.237,0.266] 4.9434473   1.351889
Effect of qsmk for p.score in (0.266,0.299] 5.5667116   1.404462
Effect of qsmk for p.score in (0.299,0.344] 2.3506362   1.312138
Effect of qsmk for p.score in (0.344,0.417] 0.9952724   1.278424
Effect of qsmk for p.score in (0.417,0.777] 3.1047566   1.222830
                                                 2.5 
Effect of qsmk for p.score in [0.051,0.124] -4.4865341 4.901220
Effect of qsmk for p.score in (0.124,0.16]   1.5255755 8.302274
Effect of qsmk for p.score in (0.16,0.189]   1.5660526 7.830214
Effect of qsmk for p.score in (0.189,0.212] -0.5679605 5.132246
Effect of qsmk for p.score in (0.212,0.237]  1.0011753 6.652791
Effect of qsmk for p.score in (0.237,0.266]  2.2937930 7.593102
Effect of qsmk for p.score in (0.266,0.299]  2.8140173 8.319406
Effect of qsmk for p.score in (0.299,0.344] -0.2211070 4.922379
Effect of qsmk for p.score in (0.344,0.417] -1.5103923 3.500937
Effect of qsmk for p.score in (0.417,0.777]  0.7080541 5.501459
\end{example}

The above example uses the stratification option and provides the standard error and confidence interval for each strata. Similar to \verb|outcome_regression()| a p-value for each strata can also be seen by using \verb|summary()| on the model.

\subsection{Doubly-Robust Estimator}

The \verb|doubly_robust()| function trains both an outcome model and a propensity model to generate predictions for the outcome and probability of treatment respectively. By default, the model uses \verb|standardization()| and \verb|propensity_scores()| to form a doubly-robust model between standardization and IP weighting. Alternatively, any outcome and propensity models can be provided instead, but must be compatible with the predict generic function in R. Since many propensity models may not predict probabilities without additional arguments into the predict function, the predictions themselves can be given for both the outcome and propensity scores. It has the following structure:

\begin{example}
doubly_robust(data, out.mod = NULL, p.mod = NULL, f = NA, family = gaussian(),
  simple = pkg.env$simple, scores = NA, p.f = NA, p.simple = pkg.env$simple,
  p.family = binomial(), p.scores = NA, n.boot = 50, ...)
\end{example}

It has the parameters for both \verb|standardization()| and \verb|propensity_scores()| so the user has full control of the underlying models. Additionally, \verb|out.mod| and \verb|p.mod| can be specified to a model trained externally as long as they predict the outcome and probability of treatment respectively. If the generic function \verb|predict()| cannot be used with either model, the user can also specify the predictions in place of the model using the \verb|scores| and \verb|p.score| parameters respectively. 

\begin{example}
R> p.model <- propensity_scores(nhefs.nmv, simple = T)
R> p.scores <- p.model$p.scores
R> out.mod <- propensity_matching(data = nhefs.nmv)
R> db.model <- doubly_robust(out.mod = out.mod, p.scores = p.scores, data = nhefs.nmv)
R> print(db.model)
\end{example}
\begin{example}
Outcome Model

Call:
glm(formula = wt82_71 ~ qsmk * p.grp, data = data)

Predictions:
   Min. 1st Qu.  Median    Mean 3rd Qu.    Max. 
-0.7685  1.0638  2.8130  2.6383  3.4145  7.9550 

Propensity Model

Call:
NULL

Predictions:
   Min. 1st Qu.  Median    Mean 3rd Qu.    Max. 
0.04011 0.18249 0.24077 0.25734 0.32152 0.78131 

Average treatment effect of qsmk:
Estimate -  2.57336 
SE       -  0.4692396 
95
\end{example}

In the example above, a random forest model was used to generate the propensity scores; see~\citet{randomForest}. Since this model needs a specific parameter in the \verb|predict()| function and subsetting of the result to obtain propensity scores, the calculations are done externally. The outcome model used is a \verb|propensity_matching()| model with all default values. Printing the estimator, a summary of the predictions from each model is given, and the estimate for the average treatment effect along with the standard error and confidence interval. Since the propensity scores were the input for the model rather than a propensity model, the call in the output is empty. Like the models in the package, this estimator is also compatable with generic functions \verb|summary()| and \verb|predict()| The \verb|summary()| function will print a summary of both models. If there are not two models such as this example, the output will be empty for that part of the estimator. The \verb|predict()| function used on the estimator will simply return the estimated average treatment effect.

\subsection{Standard Instrumental Variable Estimator}

The \verb|iv_est()| function calculates the standard IV estimand using the conditional means on a given instrumental variable. This is currently the only nonparametric function implemented in the package. While there is a method to design this estimator parametrically, that option is not implemented at this time. It has the following structure:

\begin{example}
iv_est(IV, data, n.boot = 50)
\end{example}

The instrumental variable can be binary or continuous. A string of the instrumental variable name should be assigned to the \verb|IV| parameter. 

\begin{example}
R> nhefs.iv <- nhefs[which(!is.na(nhefs$wt82) & !is.na(nhefs$price82)),]
R> nhefs.iv$highprice <- as.factor(ifelse(nhefs.iv$price82>=1.5, 1, 0))
R> nhefs.iv$qsmk <- as.factor(nhefs.iv$qsmk)
R> iv_est("highprice", nhefs.iv, n.boot = 0)
\end{example}
\begin{example}
      ATE SE 2.5 
1 2.39627 NA    NA     NA
\end{example}

In this example, a new variable was created that was an instrumental variable for the data. The parameter \verb|n.boot| was set to zero to show that bootstrapping would not occur unless the parameter is a positive integer. A data frame is returned with the estimate standard error and confidence interval.

\subsection{Single parameter G-estimation}

The \verb|gestimation()| function trains either a linear mean model for closed form estimations or a grid search. By default, the function will perform a grid search and require values for the search. It has the following structure:

\begin{example}
gestimation(data, grid, ids = list(), f = NA, family = binomial(), simple = pkg.env$simple,
  p.f = NA, p.simple = pkg.env$simple, p.family = binomial(), p.scores = NA, SW = TRUE,
  n.boot = 100, type = "one.grid", ...)
\end{example}

The formula and propensity parameters have the same meaning as previous functions in relation to an underlying propensity model. In a grid search, a \verb|geeglm()| model is used for estimation. The ids parameter is to identify the clusters of the \verb|geeglm()| model and should be the same as the number of samples in the data. The \verb|gestimation()| function will use the row names by default. Additionally, values for the search grid are required. These can be any list of numeric values that are within the estimated range for the treatment effect. The 95\% convidence interval for this method is constructed using the minimum and maximum values for $\beta$ that are statistically significant ($p < 0.05$). In a linear mean model, a closed form g-estimation will be performed and the confidence interval will be constructed by bootstrapping with the \verb|n.boot| parameter.

\begin{example}
    R> gest.model <- gestimation(nhefs.nmv, type = "one.linear")
    R> print(gest.model)
\end{example}
\begin{example}
    Call:  glm(formula = f, family = family, data = data, weights = weights)

Coefficients:
                       (Intercept)                                sex1  
                        -2.3321457                          -0.6143474  
                             race1                          education2  
                        -0.7963585                           0.0973526  
                        education3                          education4  
                         0.1996545                           0.2953593  
                        education5                           exercise1  
                         0.6016503                           0.4167123  
                         exercise2                             active1  
                         0.3579257                           0.1113038  
                           active2                                 age  
                         0.1361076                           0.1408696  
                      I(age * age)                      smokeintensity  
                        -0.0009840                          -0.0718707  
I(smokeintensity * smokeintensity)                            smokeyrs  
                         0.0008645                          -0.0847584  
            I(smokeyrs * smokeyrs)                                wt71  
                         0.0009799                          -0.0243495  
                    I(wt71 * wt71)  
                         0.0001830  

Degrees of Freedom: 1565 Total (i.e. Null);  1547 Residual
Null Deviance:	    1981 
Residual Deviance: 1853 	AIC: 1968

Average treatment effect of qsmk:
Estimate -  3.467472 
SE       -  0.5520542 
95
\end{example}

\section{Results}

\begin{figure}[ht]
\centering
\includegraphics[width=300pt]{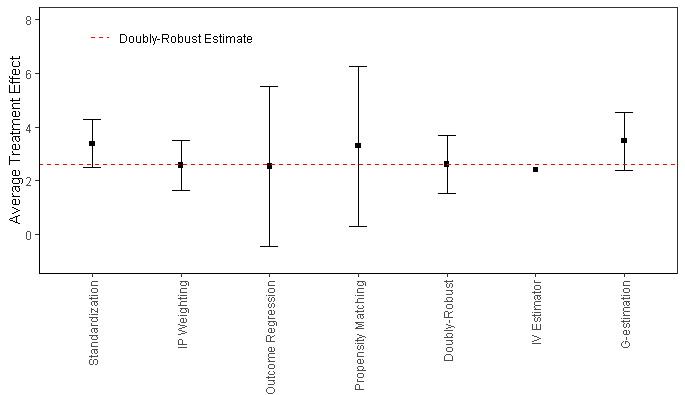}
\caption{ATE estimates from each estimator in the \CRANpkg{CausalModels} package. The error bars for each point estimate represent the 95\% Wald confidence interval from the estimator.}
\label{fig:2}
\end{figure}

After generating an estimate from each model in the package, we compare each model estimate including its 95\% confidence interval; see Figure~\ref{fig:2}. Since doubly-robust models only require one of the two models to be correctly specified, we use this as our baseline comparison among all the estimators.

IP weighting, outcome regression and the instrumental variable estimator are all in agreement with the double-robust model that the causal effect of quitting smoking on weight is approximately a weight loss of 2.4 kilograms. For this reason, we can claim all of these models are correctly specified and give an unbiased estimate of the causal effect. We can also infer standardization, propensity matching, and g-estimation seem to be wrongly specified estimating approximately 3.5 kilograms for this problem, and we should not assume this accurately estimates the true causal effect.

\section{Conclusion}
Research in causal inference requires strict attention to possible biases and correct model specification. \packageName\ offers a single source for an array of structural models for causal inference established in \citet{hernan2020causal} with a simple modeling pipeline for easy model comparison. The package focuses on abstracting repetitive steps in causal analysis while encouraging the user to adjust for bias with correct model specification. This is done while also allowing the user to adjust any model to the same extent as they would using base R. We look forward to growing the package to support more methods and a wider array of variables in the future.

\section{Availability}
This paper presented introductory concepts to causal inference and fundamental modeling methods included in the \packageName\ package available from the Comprehensive R Archive Network (CRAN) at \url{https://CRAN.R-project.org/package=CausalModels}. 

\newpage

\bibliography{CausalModels}

\address{Joshua Wolff Anderson\\
  Department of Intelligent Systems\\
  University of Pittsburgh\\
  Pittsburgh PA, USA\\
  \email{jwa45@pitt.edu}}

\address{Cyril Rakovski\\
  Department of Mathematics\\
  Chapman University\\
  Orange CA, USA\\
  \email{rakovski@chapman.edu}}

\end{article}

\end{document}